\documentclass[12pt]{article}

\usepackage{float}

\usepackage[T1]{fontenc}
\usepackage[date=year,style=numeric,giveninits=true,sorting=none,maxnames=20]{biblatex}
\AtEveryBibitem{\clearlist{language} \clearlist{location} }
\renewbibmacro{in:}{}

\usepackage[dvipdfm,margin=1.1in]{geometry}
\usepackage{amsfonts,mathtools,amsthm,bm,amssymb}
\usepackage{subfiles}
\usepackage{subcaption}
\usepackage{graphicx}
\usepackage{authblk}
\DeclareMathOperator{\sgn}{sgn}
\usepackage{color}
\usepackage[dvipdfmx]{hyperref}
\hypersetup{
setpagesize=false,
 colorlinks=true,%
 linkcolor=magenta,
 citecolor=blue,
}
\newtheorem{theorem}{Theorem}[section]

\newtheorem{corollary}[theorem]{Corollary}
\newtheorem{lemma}[theorem]{Lemma}
\newtheorem{proposition}[theorem]{Proposition}

\newtheorem{assumption}[theorem]{Assumption}

\title{Localization of space-inhomogeneous \\ three-state quantum walks}
\author[1]{Chusei Kiumi}
\affil[1]{\footnotesize Mathematical Science Unit, Graduate School of Engineering Science, Yokohama National University, Hodogaya, Yokohama, 240-8501, Japan}
\date{\empty}

\begin{document}
\maketitle
\vspace{-1.2cm}
\begin{abstract}
Mathematical analysis on the existence of eigenvalues is essential because it is equivalent to the occurrence of localization, which is an exceptionally crucial property of quantum walks. We construct the method for the eigenvalue problem via the transfer matrix for space-inhomogeneous $n$-state quantum walks in one dimension with $n-2$ self-loops, which is an extension of the technique in a previous study (Quantum Inf. Process 20(5), 2021). This method reveals the necessary and sufficient condition for the eigenvalue problem of a two-phase three-state quantum walk with one defect whose time evolution varies in the negative part, positive part, and at the origin. 
\end{abstract}

\section{Introduction}
The research on quantum walks (QWs) began in the early 2000s \cite{Ambainis2001-os,Konno2002-sh}, and QWs play important roles in various fields, and a variety of QW models have been analyzed theoretically and numerically. 
This paper focuses on the mathematical analysis of discrete-time three-state QWs on the integer lattice, studied intensively by \cite{Inui2005-ry,Wang2015-cl,Endo2019-ie,Rajendran2018-ss,Wang2015-oy,Falcao2021-pb,Saha2021-of}.
Three-state QWs have an interesting property called localization, where the probability of finding the particle around the initial position remains positive in the long-time limit. Localization on multi-state (three or more states) QWs is actively researched by previous studies \cite{Inui2005-fr,Stefanak2014-jh,Falkner2014-bt,Li2015-il,Machida2015-oa,Xu2016-dy,Kiumi2021-mg}. In the research of multi-state QWs, the Grover walk, whose time evolution is defined by the Grover matrix as a coin matrix, often plays an essential role. The name comes from Grover’s search algorithm \cite{Grover1996-de}. Two-phase two-state QWs with one defect whose time evolution varies in the negative part, positive part, and at the origin are also investigated intensively \cite{Konno2010-fx,Cantero2012-yk,Endo2015-db,Endo2014-bu,Endo2015-cy,Wojcik2012-kr,Endo2020-or,Kiumi2021-yg}. This model contains one-defect QWs where the walker at the origin behaves differently and two-phase QWs where the walker behaves differently in each negative and non-negative part. Localization on one-defect Grover walk is applied for quantum search algorithms \cite{Ambainis2005-ha,Childs2004-xa,Shenvi2003-jw}, which are expected to generalize Grover’s algorithm and speed up the search algorithms on general graphs. Also, localization on two-phase QW is related to the research of topological insulators \cite{Kitagawa2010-su,Endo2015-db}.

The QW model exhibits localization if and only if there exists an eigenvalue of the time evolution operator, and the amount of localization is deeply related to its corresponding eigenvector and the initial state of the model \cite{Segawa2016-qu}. Solving the eigenvalue problem via the transfer matrix was constructed for two-phase two-state QWs with one defect in \cite{Kiumi2021-yg} and for more general space-inhomogeneous QWs in \cite{Kiumi2021-dp}. The transfer matrix is also used in \cite{Kawai2017-fn,Kawai2018-ry,Danaci2019-qw}. In this paper, we extend the transfer matrix method to $n$-state QWs with $n-2$ self-loops. Furthermore, we apply the techniques to two-phase three-state QWs with one defect defined by the generalized Grover matrix.

The rest of this paper is organized as follows. In Section \ref{sec:2}, we define $n$-state QWs with $n-2$ self-loops, which is an extension of three-state QWs on the integer lattice. Then, we give the transfer matrix in a general way and construct methods for the eigenvalue problem.
Theorem \ref{Theorem Ker} is the main theorem, which gives a necessary and sufficient condition for the eigenvalue problem.  Section \ref{sec:3} focuses on the concrete calculation of eigenvalues on one-defect and two-phase three-state QWs with generalized Grover coin matrices. We also show some figures indicating eigenvalues of the time evolution operators and their corresponding probability distributions in this section.

\section{Definitions and Method}
\label{sec:2}
\subsection{Multi-state quantum walks on the integer lattice}
Firstly, we introduce $n$-state QWs with $n-2$ self-loops on the integer lattice $\mathbb{Z}$.
Let $\mathcal{H}$ be a Hilbert space defined by
\[
\mathcal{H}=\ell^2(\mathbb{Z} ; \mathbb{C}^n) =
\left\{
\Psi : \mathbb{Z} \to \mathbb{C}^n\ \middle\vert\ \sum_{x\in\mathbb{Z}}\|\Psi(x)\|_{\mathbb{C}^n}^2 < \infty
\right\},
\]
where $n \geq 3$ and $\mathbb{C}$ denotes the set of complex numbers. We write $n$-state quantum state $\Psi:\mathbb{Z}\rightarrow \mathbb{C}^n$ as below:
\[
\Psi (x)=\left[\begin{array}{ c }
\Psi _{1} (x)\\
\Psi _{2} (x)\\
\vdots \\
\Psi _{n} (x)
\end{array}\right].
\]
Let $\{C_x\}_{x\in\mathbb{Z}}$ be a sequence of $n\times n$ unitary matrices, which is written as below:
\[
C_{x} =e^{i\Delta _{x}}\left[\begin{array}{ c c c c }
a_{x}^{( 1,1)} & a_{x}^{( 1,2)} & \dotsc  & a_{x}^{( 1,n)}\\
a_{x}^{( 2,1)} & a_{x}^{( 2,2)} & \dotsc  & a_{x}^{( 2,n)}\\
\vdots  & \vdots  & \ddots  & \vdots \\
a_{x}^{( n,1)} & a_{x}^{( n,2)} & \dotsc  & a_{x}^{( n,n)}
\end{array}\right].\]
 where $\Delta_x \in [0,2\pi),\  a^{(j,k)}_x\in\mathbb{C},(1\leq j, k\leq n)$ and $|a^{(k,k)}_x|\neq 1 \ (2\leq k\leq n-1)$. Here we define $C_x$ with additional phases $\Delta_x$ for the simplification of the discussion in Subsection \ref{subsec:transfer}.
Then the coin operator $C$ on $\mathcal{H}$ is given as 
\[(C\Psi)=C_x\Psi(x).\]
 The shift operator $S$ is also an operator on $\mathcal{H}$, which shifts $\Psi_1(x)$ and $\Psi_n(x)$ to $\Psi_1(x-1)$ and $\Psi_n(x+1)$, respectively and does not move $\Psi_k(x)$ for $2\leq k \leq n-1$.

\[
(S\Psi )(x)=\left[\begin{array}{ c }
\Psi _{1} (x+1)\\
\Psi _{2} (x)\\
\vdots \\
\Psi _{k} (x)\\
\vdots \\
\Psi _{n-1} (x)\\
\Psi _{n} (x-1)
\end{array}\right] ,\qquad 2\leq k\leq n-1.\ \]
Then the time evolution operator is given as  
\[
U=SC.
\]
We treat the model whose coin matrices satisfy
\[
C_x =
\begin{cases}
C_\infty ,\quad & x\in [ x_{+} ,\infty ),
\\
C_{-\infty} ,\quad & x\in ( -\infty ,x_{-}].
\end{cases}
\]
where $x_+>0,\ x_-<0$. For initial state $\Psi_0\in\mathcal{H}\ (\|\Psi_0\|_{\mathcal{H}}^2=1)$, the finding probability of a walker in position $x$ at time $t\in\mathbb{Z}_{\geq 0}$ is defined by 
\[\mu_t^{(\Psi_0)}(x)=\|(U^t\Psi_0)(x)\|_{\mathbb{C}^n}^2,\] 
where $\mathbb{Z}_{\geq 0}$ is the set of non-negative integers. We say that the QW exhibits localization if there exists a position $x_0\in\mathbb{Z}$ and an initial state $\Psi_0\in\mathcal{H}$ satisfying $\limsup_{t\to\infty}\mu^{(\Psi_0)}_t(x_0)>0$. It is known that the QW exhibits localization if and only if there exists an eigenvalue of $U$ \cite{Segawa2016-qu}, that is, there exists $\lambda\in[0, 2\pi)$ and $\Psi\in\mathcal{H}\setminus\{\mathbf{0}\}$  such that
\begin{align*}
	U\Psi=e^{i\lambda}\Psi.
	\end{align*}
Let $\sigma_p(U)$ denotes the set of eigenvalues of $U$, henceforward.

\subsection{Eigenvalue problem and transfer matrix}
\label{subsec:transfer}
The method to solve the eigenvalue problem of space-inhomogeneous two-state QWs with the transfer matrix was introduced in \cite{Kiumi2021-yg,Kiumi2021-dp}. This subsection shows that the transfer matrix method can also be applied to $n$-state QWs with $n-2$ self-loops. Firstly, $U\Psi =e^{i\lambda }\Psi$ is equivalent that $\Psi\in\mathcal{H}$ satisfies followings for all $x\in\mathbb{Z}$:
\[
e^{i( \lambda -\Delta _{x})} \Psi _{1} (x-1)=\sum _{i=1}^{n} a_{x}^{( 1,i)} \Psi _{i} (x),\quad
e^{i( \lambda -\Delta _{x})} \Psi _{n} (x+1)=\sum _{i=1}^{n} a_{x}^{( n,i)} \Psi _{i} (x),
\]
and for $2\leq k\leq n-1$
\[
e^{i( \lambda -\Delta _{x})} \Psi _{k} (x)=\sum _{i=1}^{n} a_{x}^{( k,i)} \Psi _{i} (x)\Longleftrightarrow \Psi _{k} (x)=\frac{\sum _{i=1,i\neq k}^{n} a_{x}^{( k,i)} \Psi _{i} (x)}{e^{i( \lambda -\Delta _{x})} -a_{x}^{( k,k)}}.
\]
where $\Longleftrightarrow$ denotes ``if and only if''. By repetition of substitutions, we can eliminate $\Psi_k(x)\ (2\leq k\leq n-1)$ from this system of equations, and this can be converted to the following equivalent system of equations:
\begin{align}
\label{eq1}
 & e^{i( \lambda -\Delta _{x})} \Psi _{1} (x-1)=A_{x}( \lambda ) \Psi _{1} (x)+B_{x}( \lambda ) \Psi _{n} (x),\\
 \label{eq2}
 & e^{i( \lambda -\Delta _{x})} \Psi _{n} (x+1)=C_{x}( \lambda ) \Psi _{1} (x)+D_{x}( \lambda ) \Psi _{n} (x),\\
 \label{eq3}
 & \Psi _{k} (x)=E_{k,x}( \lambda ) \Psi _{1} (x)+F_{k,x}( \lambda ) \Psi _{n} (x),
\end{align}
where $A_x(\lambda),B_x(\lambda),C_x(\lambda),D_x(\lambda),E_{k,x}(\lambda),F_{k,x}(\lambda)$ are $\mathbb{C}$-valued function and their absolute values are finite real numbers. When $\displaystyle n=3,$ these values become
\begin{align*}
 & A_{x}( \lambda ) =a_{x}^{( 1,1)} +\frac{a_{x}^{( 1,2)} a_{x}^{( 2,1)}}{e^{i( \lambda -\Delta _{x})} -a_{x}^{( 2,2)}} ,\ B_{x}( \lambda ) =a_{x}^{( 1,3)} +\frac{a_{x}^{( 1,2)} a_{x}^{( 2,3)}}{e^{i( \lambda -\Delta _{x})} -a_{x}^{( 2,2)}} ,\\
 & C_{x}( \lambda ) =a_{x}^{( 3,1)} +\frac{a_{x}^{( 3,2)} a_{x}^{( 2,1)}}{e^{i( \lambda -\Delta _{x})} -a_{x}^{( 2,2)}} ,\ D_{x}( \lambda ) =a_{x}^{( 3,3)} +\frac{a_{x}^{( 3,2)} a_{x}^{( 2,3)}}{e^{i( \lambda -\Delta _{x})} -a_{x}^{( 2,2)}} ,\\
 & E_{2,x}( \lambda ) =\frac{a_{x}^{( 2,1)}}{e^{i( \lambda -\Delta _{x})} -a_{x}^{( 2,2)}} ,\ F_{2,x}( \lambda ) =\frac{a_{x}^{( 2,3)}}{e^{i( \lambda -\Delta _{x})} -a_{x}^{( 2,2)}} .
\end{align*}
 Note that $\Psi:\mathbb{Z}\rightarrow\mathbb{C}^n$, where $\Psi$ does not necessarily satisfy $\|\sum_{x\in\mathbb{Z}}\Psi(x)\|_{\mathbb{C}^n}^2<\infty$ but satisfies (\ref{eq1}), (\ref{eq2}), (\ref{eq3}) is a generalized eigenvector of $U$, which is the stationary measure of QWs studied in \cite{Wang2015-oy,Kawai2017-fn,Kawai2018-ry,Endo2019-ie}. Here, we define transfer matrices $T_x(\lambda)$ by
\[
T_x(\lambda)=\frac{1}{A_{x}( \lambda )}\begin{bmatrix}
e^{i( \lambda -\Delta _{x})} & -B_{x}( \lambda )\\
C_{x}( \lambda ) & -e^{-i( \lambda -\Delta _{x})}( B_{x}( \lambda ) C_{x}( \lambda ) -A_{x}( \lambda ) D_{x}( \lambda ))
\end{bmatrix},
\]
then equations (\ref{eq1}), (\ref{eq2}) can be written as 
\[
\begin{bmatrix}
\Psi _{1} (x)\\
\Psi _{n} (x+1)
\end{bmatrix} =T_x(\lambda)\begin{bmatrix}
\Psi _{1} (x-1)\\
\Psi _{n} (x)
\end{bmatrix}.
\]
Note that, when $A_x(\lambda)=0$, we cannot construct a transfer matrix. Therefore, we have to treat the case $A_x(\lambda)=0$ separately. For simplification, we write $T_x(\lambda)$ as $T_x$ henceforward. Let $\lambda\in[0,2\pi)$ satisfying $A_x(\lambda)\neq0$ for all $x\in\mathbb{Z}$ and $\varphi \in\mathbb{C}^2$, we define $\tilde{\Psi}:\mathbb{Z}\rightarrow\mathbb{C}^2$ as follows: 
	\begin{align}
	\label{cor:tilde_psi}
\tilde{\Psi } (x) & =\left\{\begin{array}{ l l }
T_{x-1} T_{x-2} \cdots T_{1} T_{0} \varphi , & x >0,\\
\varphi , & x=0,\\
T^{-1}_{x} T^{-1}_{x+1} \cdots T^{-1}_{-2} T^{-1}_{-1} \varphi , & x< 0.
\end{array}\right. \nonumber\\[+8pt]
 & =\begin{cases}
T_{\infty}^{x-x_{+}} T_{+} \varphi,  & x_{+} \leq x,\\
T_{x-1} \cdots T_{0} \varphi,  & 0< x< x_{+},\\
\varphi,  & x=0,\\
T^{-1}_{x} \cdots T^{-1}_{-1} \varphi,  & x_{-} < x< 0,\\
T^{x-x_{-}}_{-\infty } T_{-} \varphi,  & x \leq x_{-}.
\end{cases}
\end{align}
where $T_{+}=T_{x_{+} -1} \cdots T_{0},\  T_{-} = T^{-1}_{x_{-}} \cdots T^{-1}_{-1}$ and $T_{\pm \infty} = T_{x_\pm}$. 
Let $V_{\lambda}$ be a set of generalized eigenvectors and $W_{\lambda}$ be a set of reduced vectors $\tilde\Psi$ defined by (\ref{cor:tilde_psi}):
\begin{align*}
V_{\lambda}&=\left\{\Psi:\mathbb{Z}\rightarrow\mathbb{C}^n\ \middle|\  \Psi \text{ satisfies } (\ref{eq1}),(\ref{eq2}), (\ref{eq3})\right\},
\\
W_{\lambda}&=\left\{\tilde\Psi:\mathbb{Z}\rightarrow\mathbb{C}^2\ \middle|\  \tilde\Psi(x)\text{ is given by }  (\ref{cor:tilde_psi}),\  \varphi\in\mathbb{C}^2\right\},
\end{align*}
 for $\lambda\in[0,2\pi)$ satisfying $A_x(\lambda)\neq0$ for all $x\in\mathbb{Z}$. We define bijective map $\iota:V_{\lambda}\rightarrow W_{\lambda}$ by
\[
(\iota \Psi)(x)=\left[\begin{array}{c}
\Psi_{1}(x-1) \\
\Psi_{n}(x)
\end{array}\right].
\]
Here, the inverse of $\iota$ is given as
\begin{align}
\label{eq:inverse}
\left( \iota ^{-1}\tilde{\Psi }\right)( x) \begin{bmatrix}
\tilde{\Psi }_{1}( x+1)\\
E_{2,x}( \lambda )\tilde{\Psi }_{1} (x+1)+F_{2,x}( \lambda )\tilde{\Psi }_{2} (x)\\
\vdots \\
E_{k,x}( \lambda )\tilde{\Psi }_{1} (x+1)+F_{k,x}( \lambda )\tilde{\Psi }_{2} (x)\\
\vdots \\
E_{n-1,x}( \lambda )\tilde{\Psi }_{1} (x+1)+F_{n-1,x}( \lambda )\tilde{\Psi }_{2} (x)\\
\tilde{\Psi }_{2}( x)
\end{bmatrix}
\end{align}
for $\tilde\Psi\in W_{\lambda}$. Thus,
$\Psi\in V_\lambda$ if and only if there exists $\tilde\Psi\in W_\lambda$ such that $\Psi=\iota^{-1}\tilde\Psi$. From the definition of $\iota$, we can also say that $\iota^{-1}\tilde\Psi\in\mathcal{H}\setminus\{\mathbf{0}\}$
if and only if $\tilde\Psi\in\ell^2(\mathbb{Z};\mathbb{C}^2)\setminus\{\mathbf{0}\}$. Therefore, we get the following corollary.
\begin{corollary}
\label{cor:ell2}
Let $\lambda\in [0,2\pi)$ satisfying $A_x(\lambda)\neq0$ for all $x$, $e^{i\lambda}\in\sigma_p(U)$ if and only if there exists $\tilde\Psi\in W_{\lambda} \setminus\{\mathbf{0}\}$ such that $\tilde\Psi\in\ell^2(\mathbb{Z};\mathbb{C}^2)$, and associated eigenvector of $e^{i\lambda}$ becomes $\iota^{-1}\tilde\Psi$.
\end{corollary}

\vspace{5mm}

In this paper, since we focus on the eigenvalue problem for generalized three-state Grover walks defined by coin matrices (\ref{eq:grover}) , we consider the following assumptions:
\begin{assumption}
\label{assumption}
$\lambda\in [0,2\pi)$ satisfies following conditions:
\begin{align*}
 & 1.\ A_{x}( \lambda ) \neq 0,\ for\ all\ x\in \mathbb{Z} ,\\
 & 2.\ \det( T_{\pm \infty }) =\frac{D_{\pm \infty }( \lambda )}{A_{\pm \infty }( \lambda )} =1,\\
 & 3.\ \mathrm{tr}( T_{\pm \infty }) \in \mathbb{R},
\end{align*}
\end{assumption}
We define sign function for real numbers $r$ as follows:
\[
\sgn( r) =\begin{cases}
1, & r >0,\\
0, & r=0,\\
-1, & r< 0.
\end{cases}
\]
The pair of eigenvalues of $T_{\pm \infty }$ can be written as $\zeta _{\pm \infty}^{ >},\zeta _{\pm \infty}^{ <}$ defined by 
\[
  \zeta _{\pm \infty}^{ >} =\frac{\mathrm{tr}( T_{\pm \infty}) +\sgn( \mathrm{tr}( T_{\pm \infty}))\sqrt{\mathrm{tr}( T_{\pm \infty })^{2} -4}}{2},
\]
\[
\zeta _{\pm \infty}^{< } =\frac{\mathrm{tr}( T_{\pm \infty}) -\sgn( \mathrm{tr}( T_{\pm \infty}))\sqrt{\mathrm{tr}( T_{\pm \infty })^{2} -4}}{2},
\]
where $|\zeta _{\pm \infty}^{ >}|\geq 1$  and $|\zeta _{\pm \infty}^{ <}|\leq 1$ since $|\zeta^{>}_{\pm \infty}||\zeta^{<} _{\pm \infty}|=|\det(T_{\pm\infty})|=1$ holds. Hence, we have the main theorem.
\begin{theorem}
\label{Theorem Ker}
Under the Assumption \ref{assumption}, $e^{i\lambda} \in \sigma _{p}(U) $ if and only if following two conditions hold:
\begin{align*}
    &1.\ \mathrm{tr}( T_{\pm \infty })^{2} -4>0,\\
    &2.\ \ker\left(\left( T_{\infty} -\zeta ^{<}_{\infty }\right) T_{+}\right) \cap \ker\left(\left( T_{-\infty} -\zeta ^{>}_{-\infty }\right) T_{-}\right)\neq \{\mathbf{0}\}.
\end{align*}
\begin{proof}
From Corollary \ref{cor:ell2}, $e^{i\lambda} \in \sigma _{p}(U)$ if and only if there exists $\tilde\Psi\in W_{\lambda} \setminus\{\mathbf{0}\}$ such that $\sum_{x\in\mathbb{Z}}\|\tilde\Psi(x)\|_{\mathbb{C}^2}^2<\infty$. Firstly, when $\mathrm{tr}( T_{\pm \infty })^{2} -4\leq 0$, both $|\zeta ^<_{\pm\infty}|$ and $|\zeta^>_{\pm\infty}|$ become 1. Since  $\tilde\Psi(x)$ is given as (\ref{cor:tilde_psi}), $\sum_{x\in\mathbb{Z}}\|\tilde\Psi(x)\|_{\mathbb{C}^2}^2=\infty$ for all $\tilde\Psi\in W_{\lambda}\setminus\{\mathbf{0}\}$. Therefore, $\mathrm{tr}( T_{\pm \infty })^{2} -4> 0$ is a necessary condition for $e^{i\lambda} \in \sigma _{p}(U)$. Secondly, if $\mathrm{tr}( T_{\pm \infty })^{2} -4> 0$, then $|\zeta^>_{\pm\infty}|>1$ and $|\zeta^<_{\pm\infty}|<1$ hold. Since $\tilde\Psi\in W_{\lambda}\setminus\{\mathbf{0}\}$ is expressed by $\varphi \in \mathbb{C}^2\setminus \{\mathbf{0}\}$ and transfer matrices, there exists $\tilde\Psi\in W_{\lambda}\setminus\{\mathbf{0}\}$ such that $\sum_{x\in\mathbb{Z}}\|\tilde\Psi(x)\|_{\mathbb{C}^2}^2<\infty$ if and only if there exists $\varphi \in \mathbb{C}^2\setminus \{\mathbf{0}\}$ such that $T_{+}\varphi\in \ker\left( T_{\infty} -\zeta ^{<}_{\infty }\right),\ T_{-}\varphi \in \ker\left( T_{-\infty} -\zeta ^{>}_{-\infty }\right)$, that is, $\varphi\in \ker\left(\left( T_{\infty} -\zeta ^{<}_{\infty }\right) T_{+}\right) \cap \ker\left(\left( T_{-\infty} -\zeta ^{>}_{-\infty }\right) T_{-}\right).$ From these discussions, we have proved the statement.
\end{proof}
\end{theorem}

From Theorem \ref{Theorem Ker}, when $e^{i\lambda} \in \sigma _{p}( U)$, $\Psi\in\ker(U-e^{i\lambda})\setminus\{\mathbf{0}\}$ is given as $\Psi=\iota^{-1}\tilde\Psi$ where $\tilde\Psi\in W_\lambda\setminus\{\mathbf{0}\}$  is expressed as
\begin{align*}
\tilde{\Psi } (x)  =\begin{cases}
(\zeta ^{<}_{\infty })^{x-x_{+}} T_{+} \varphi,  & x_{+} \leq x,\\
T_{x-1} \cdots T_{0} \varphi,  & 0< x< x_{+},\\
\varphi,  & x=0,\\
T^{-1}_{x} \cdots T^{-1}_{-1} \varphi,  & x_{-} < x< 0,\\
(\zeta ^{>}_{-\infty })^{x-x_{-}} T_{-} \varphi,  & x \leq x_{-}.
\end{cases}
\end{align*}
with $\varphi\in\ker\left(\left( T_{\infty} -\zeta ^{<}_{\infty }\right) T_{+}\right) \cap \ker\left(\left( T_{-\infty} -\zeta ^{>}_{-\infty }\right) T_{-}\right)\setminus\{\mathbf{0}\}.$ From \cite{Kiumi2021-yg,Kiumi2021-dp}, we can also say that $\dim\ker(U-e^{i\lambda})=1$ under the Assumption \ref{assumption}.

\section{Eigenvalues of three-state Grover walks}
\label{sec:3} In this section, we focus on the following generalized Grover matrices as the coin matrix, which is the coin matrix studied in \cite{Machida2015-oa} with an additional phase $\Delta_x$.
\begin{align}
\label{eq:grover}
C_{x} =e^{i\Delta _{x}}\left[\begin{array}{ c c c }
-\dfrac{1+c_{x}}{2} & \dfrac{s_{x}}{\sqrt{2}} & \dfrac{1-c_{x}}{2}\\
\dfrac{s_{x}}{\sqrt{2}} & c_{x} & \dfrac{s_{x}}{\sqrt{2}}\\
\dfrac{1-c_{x}}{2} & \dfrac{s_{x}}{\sqrt{2}} & -\dfrac{1+c_{x}}{2}
\end{array}\right],
\end{align}
where $\displaystyle c_{x} =\cos \theta _{x} ,s_{x} =\sin \theta _{x}$ with $\displaystyle \theta_x \in [ 0,2\pi )$ and $\displaystyle \theta_x \neq 0,\pi $. Then,
\[
A_{x}( \lambda ) =D_{x}( \lambda ) =\frac{( 1+c_{x})\left( 1-e^{i( \lambda -\Delta _{x})}\right)}{2\left( e^{i( \lambda -\Delta _{x})} -c_{x}\right)} ,\ B_{x}( \lambda ) =C_{x}( \lambda ) =\frac{( 1-c_{x})\left( 1+e^{i( \lambda -\Delta _{x})}\right)}{2\left( e^{i( \lambda -\Delta _{x})} -c_{x}\right)}.\]
Transfer matrices become
\[
T_{x} =\frac{1}{( 1+c_{x})\left( 1-e^{i( \lambda -\Delta _{x})}\right)}\begin{bmatrix}
2e^{i( \lambda -\Delta _{x})}\left( e^{i( \lambda -\Delta _{x})} -c_{x}\right) & -( 1-c_{x})\left( 1+e^{i( \lambda -\Delta _{x})}\right)\\[5pt]
( 1-c_{x})\left( 1+e^{i( \lambda -\Delta _{x})}\right) & -2e^{-i( \lambda -\Delta _{x})}\left( 1-c_{x} e^{i( \lambda -\Delta _{x})}\right)
\end{bmatrix}\]
where 
\begin{align*}
  &\det( T_{x })=\frac{D_{x}( \lambda )}{A_{x}( \lambda )} =1,\\
  &\ \mathrm{tr}( T_{x}) =-\frac{2( 2\cos( \lambda -\Delta _{x}) +1-c_{x})}{( 1+c_{x})}\in\mathbb{R}.
\end{align*}
Thus, we can say that $\lambda\in(0,2\pi]$ where $\lambda\neq \Delta_x$ for all $x\in\mathbb{Z}$ satisfies Assumption \ref{assumption}.
\begin{lemma}
\label{lemma}
$e^{i\Delta_{\pm\infty}}\in\sigma_p(U).$
\begin{proof}
When $e^{i\lambda }=e^{i\Delta_{\pm\infty}}$, $\lambda$ does not satisfies Assumption \ref{assumption}, thus we consider these cases separately.
  From the discussion in Section \ref{sec:2}, $ e^{i\lambda } \in\sigma_p(U)$ is equivalent that there exists $\Psi\in\mathcal{H}\setminus\{\mathbf{0}\}$ satisfying (\ref{eq1}), (\ref{eq2}) and (\ref{eq3}). 
Considering the case $A_{\infty}( \lambda ) =0$, i.e., $ e^{i\lambda } =e^{i\Delta _{\infty}}$, $\Psi:\mathbb{Z}\rightarrow\mathbb{C}^3$ satisfies (\ref{eq1}) and (\ref{eq2}) if and only if $\Psi$ satisfies 
\begin{align}
\label{eq1:lemma}
&\Psi_{1}(x-1) =\Psi_{3}(x),\ \Psi_{1}(x) =\Psi_{3}(x+1)\ (\text{if } \Delta_x=\Delta_\infty),\\
\label{eq2:lemma}
&\left[\begin{array}{c}
\Psi_{1}(x) \\
\Psi_{3}(x+1)
\end{array}\right] =T_x(\lambda)\left[\begin{array}{c}
\Psi_{1}(x-1) \\
\Psi_{3}(x)
\end{array}\right]\ (\text{if } \Delta_x\neq\Delta_\infty).
\end{align}
Let $k\in\mathbb{C}\setminus \{0\}$ and $x_\infty\in(x_+,\infty)$. We consider $\Psi:\mathbb{Z}\rightarrow \mathbb{C}^3$ defined by
\[
\Psi (x)=\begin{cases}
 \begin{bmatrix}
k & E_{2,x}( \lambda ) k & 0
\end{bmatrix}^{t}, & x=x_{\infty } ,\\
 \begin{bmatrix}
0 & F_{2,x}( \lambda ) k & k
\end{bmatrix}^{t}
 , & x=x_{\infty }+1 ,\\
\mathbf{0} , & otherwise,
\end{cases}\]
where $t$ is a transpose operator. Then, ${\Psi}\in\mathcal{H}\setminus\{\mathbf{0}\}$ holds, and $\Psi$ satisfies condition (\ref{eq1:lemma}) (\ref{eq2:lemma}) and (\ref{eq3}). Therefore, $e^{i\Delta_\infty}\in\sigma_p(U)$.  Considering the case of  $A_{-\infty}( \lambda ) =0$ in the same way, we have $e^{i\Delta_{-\infty}}\in\sigma_p(U)$.
\end{proof}
\end{lemma}

\subsection{Two-phase quantum walks with one defect}
\label{subsec:tqwo}
Henceforward, we consider two-phase QWs with one defect ($x_+ =1,\ x_- = -1$).   
\begin{align*}
	C_x=
	\begin{cases}
	C_m,\quad &x<0,
	\\
	C_o,\quad &x=0,
	\\
	C_p,\quad &x>0.
	\end{cases}
	\end{align*}
 We write $T_x = T_j,\ \zeta_{x}^<=\zeta_{j}^<,\ \zeta_{x}^>=\zeta_{j}^>, \ A_x(\lambda)=A_j(\lambda),$ where $j=p\ (x>0)$, $=o\ (x=0)$, $=m\ (x<0)$.
    In this case, $T_+$ and $T_-$ equal $T_o$ and $T_{m}^{-1}$, respectively. We now apply Theorem \ref{Theorem Ker} to two-phase QWs with one defect case. Under Assumption \ref{assumption}, $e^{i\lambda}\in\sigma_p(U)$ if and only if there exists $\varphi\in\mathbb{C}^2\setminus\{\mathbf{0}\}$ such that
\begin{align*}
    &1.\ \cos( \lambda -\Delta _{m}) -c_{m}>0,\  \cos( \lambda -\Delta _{p}) -c_{p}>0,\\
    &2.\ T_{o}\varphi\in \ker\left( T_{p} -\zeta ^{<}_{p }\right),\ \varphi \in \ker\left( T_{m} -\zeta ^{>}_{m}\right).
\end{align*}
Here, for $e^{i\lambda } \neq \pm e^{i\Delta _{j}}$,
\begin{align}
\label{eq:ev1}
 & \ker\left( T_{j} -\zeta _{j}^{ >}\right) =\left\{k\begin{bmatrix}
t_{1} +t_{2}\\
t_{2} -t_{1} -i\sgn(\sin( \lambda -\Delta _{j}))\sqrt{4t_{1} t_{2}}
\end{bmatrix} \ \middle| \ k\in \mathbb{C}\right\} ,\\
\label{eq:ev2}
 & \ker\left( T_{j} -\zeta _{j}^{< }\right) =\left\{k\begin{bmatrix}
t_{1} +t_{2}\\
t_{2} -t_{1} +i\sgn(\sin( \lambda -\Delta _{j}))\sqrt{4t_{1} t_{2}}
\end{bmatrix} \ \middle| \ k\in \mathbb{C}\right\} .
\end{align}
where $t_{1} =1-\cos( \lambda -\Delta _{j}) ,\ t_{2} =\cos( \lambda -\Delta _{j}) -c_{j} \ ( j=m,p).$
 Also, $\Psi\in\ker(U-e^{i\lambda})\setminus\{\mathbf{0}\}$ becomes $\Psi=\iota^{-1}\tilde\Psi$ where
\[
\tilde{\Psi } (x)=\begin{cases}
T^{x-1}_{p} T_{o} \varphi,  & x >0,\\
T^{x}_{m} \varphi,  & x\leq 0.
\end{cases} 
\ 
=
\ 
\begin{cases}
\left(\zeta_{p}^{<}\right)^{x-1} T_{o} \varphi,  & x >0,\\
\left(\zeta_{m}^{>}\right)^{x} \varphi,  & x\leq 0.
\end{cases}
\]

\subsection{One-defect model}
\begin{figure}[H]
\begin{subfigure}[H]{0.49\textwidth}
\centering
\includegraphics[width=0.83\linewidth]{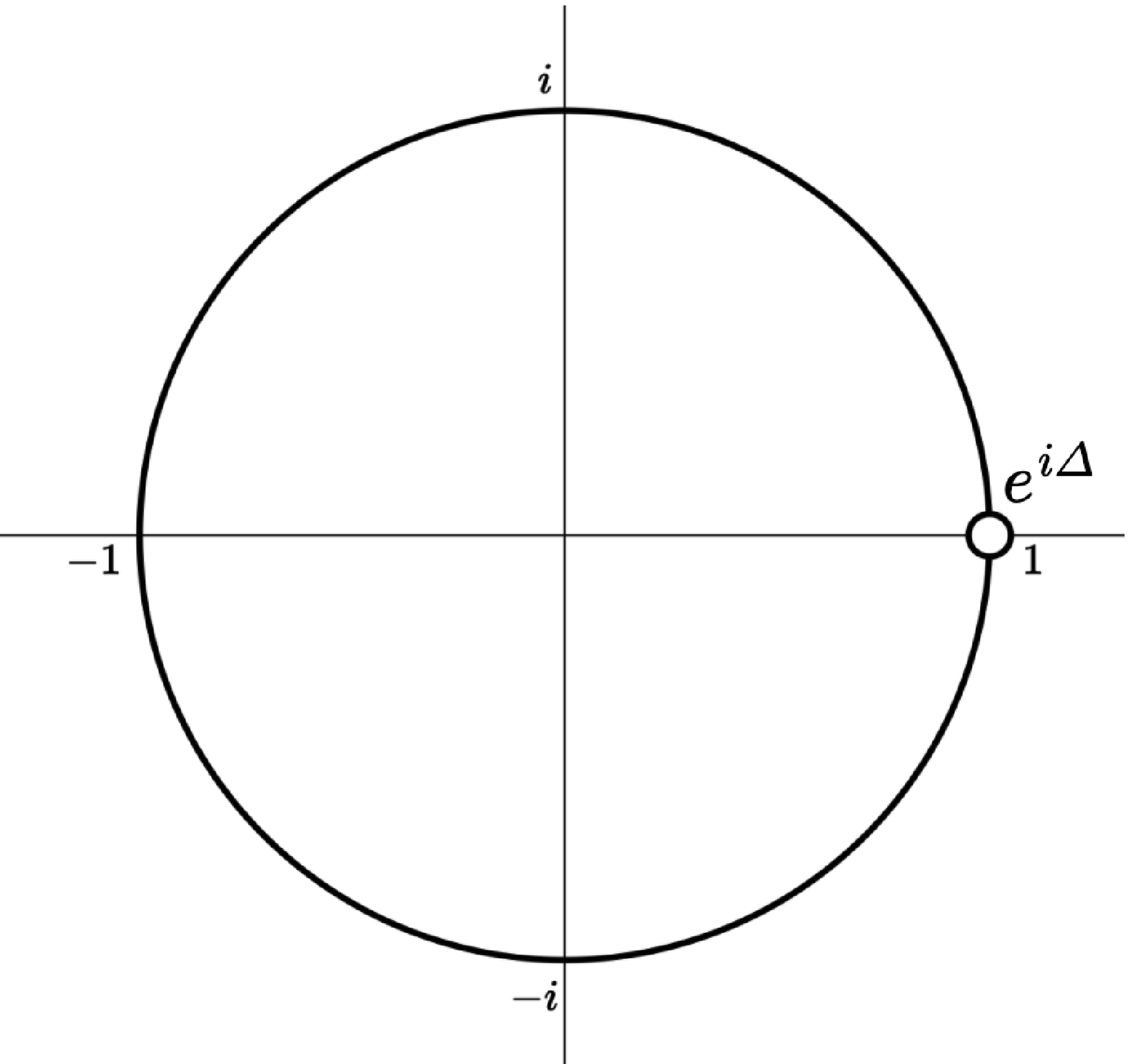} 
\caption{Eigenvalue of $U$}
\end{subfigure}
\begin{subfigure}[H]{0.49\textwidth}
\centering
\includegraphics[width=0.95\linewidth]{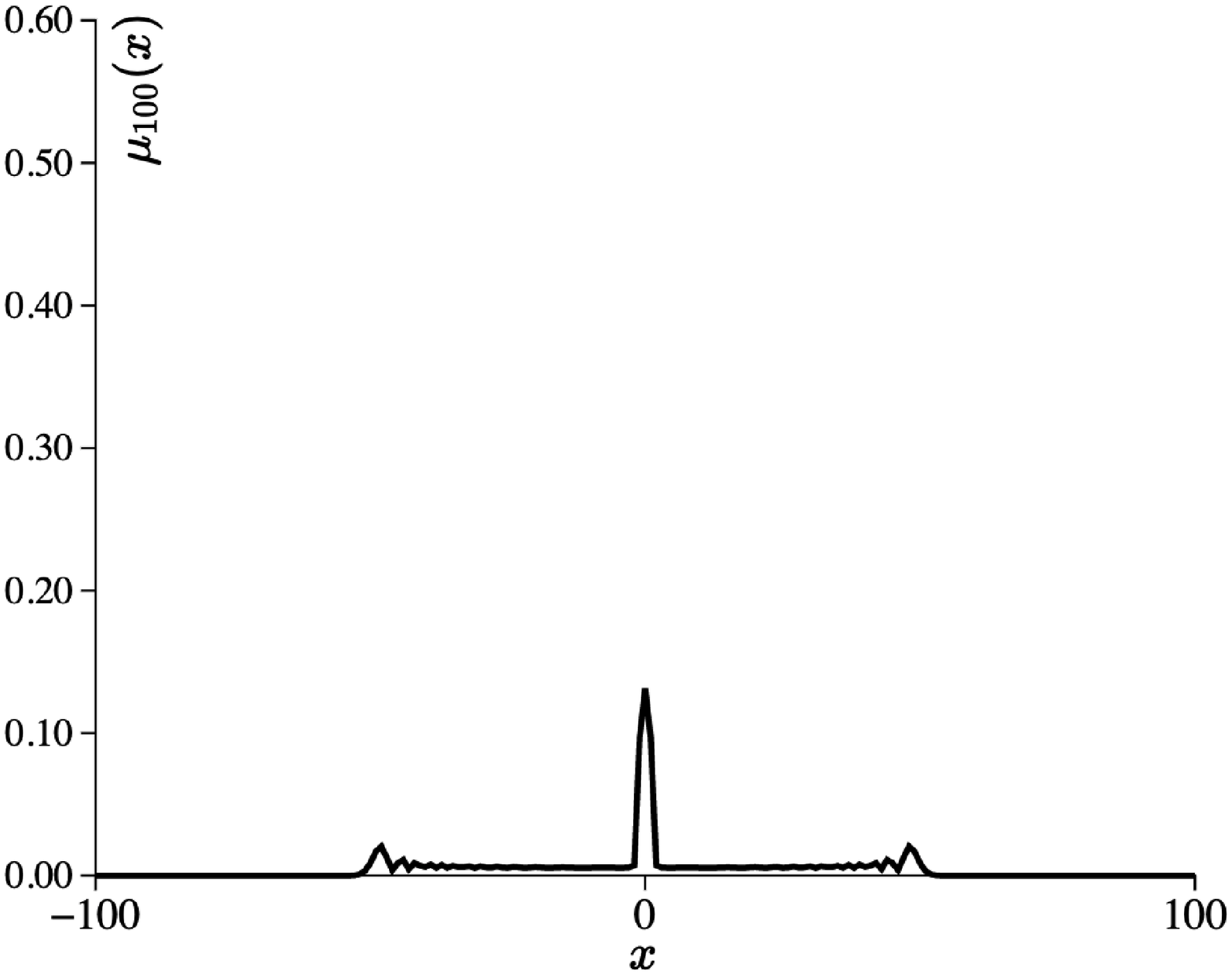}
\caption{Probability distribution}
\end{subfigure}
\caption{Example of Proposition \ref{pro:one-defect} with parameters $\displaystyle \Delta _{o} =\Delta =0,\ \theta _{o} =\frac{8}{12} ,\ \theta =-\frac{8}{12} \pi $. (a) illustrates the eigenvalue, and (b) shows the probability distribution at time $100$ with initial state $\Psi_0(0)=[\frac{1}{\sqrt{3}},\frac{i}{\sqrt{3}},\frac{1}{\sqrt{3}}]^t$ and $\Psi_0(x)=\mathbf{0}$ for $x\neq 0$.}
\label{fig:1}  
\end{figure}

\begin{figure}[H]
\begin{subfigure}[H]{0.49\textwidth}
\centering
\includegraphics[width=0.83\linewidth]{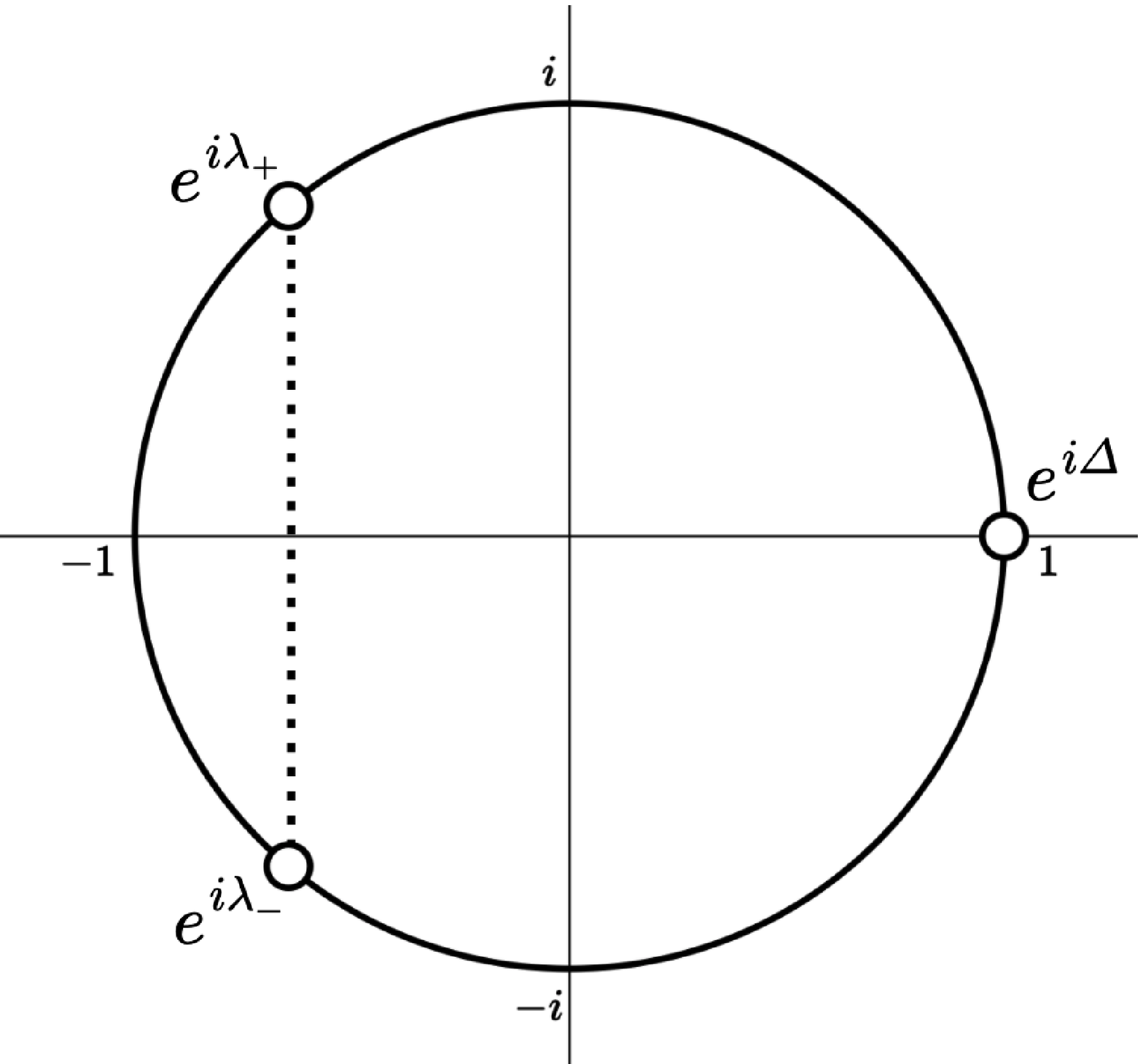} 
\caption{Eigenvalues of $U$}
\end{subfigure}
\begin{subfigure}[H]{0.49\textwidth}
\centering
\includegraphics[width=0.95\linewidth]{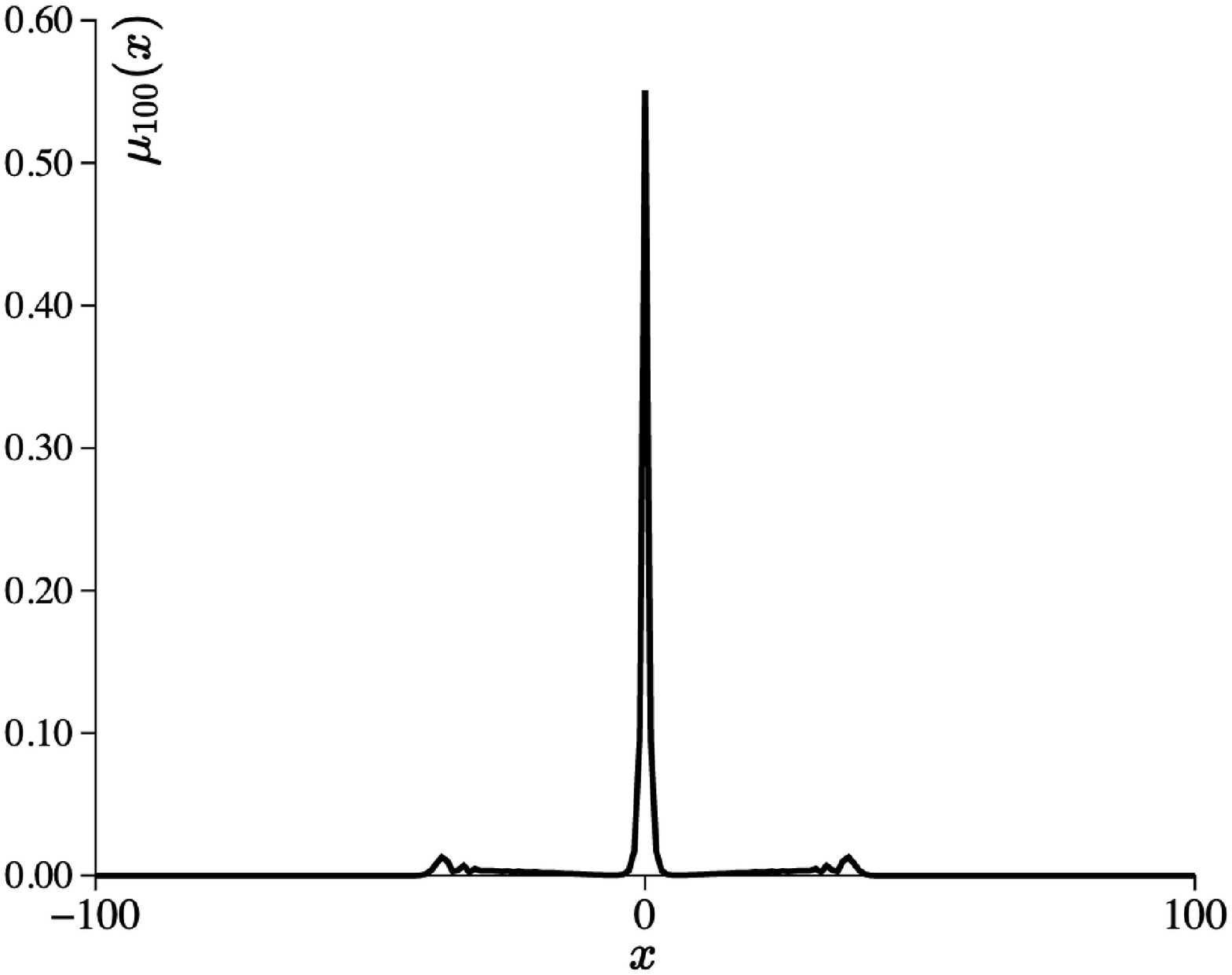}
\caption{Probability distribution}
\end{subfigure}
\caption{Example of Proposition \ref{pro:one-defect} with parameters $\displaystyle \Delta _{o} =\Delta =0,\ \theta _{o} =\frac{8}{12} \pi ,\ \theta =-\frac{9}{12} \pi $. (a) illustrates the eigenvalues, and (b) shows the probability distribution at time $100$ with initial state $\Psi_0(0)=[\frac{1}{\sqrt{3}},\frac{i}{\sqrt{3}},\frac{1}{\sqrt{3}}]^t$ and $\Psi_0(x)=\mathbf{0}$ for $x\neq 0$}
\label{fig:2}  
\end{figure}

Here, we consider the one-defect model, where $\Delta_m=\Delta_p=\Delta,\ c_m=c_p=c, \ \zeta_m^{>}=\zeta^{>},\  \zeta_p^{<}=\zeta^{<},\ T_m=T_p=T$. First, we consider $\lambda$ which does not satisfy Assumption \ref{assumption}, i.e, $e^{i\lambda}= e^{i\Delta},e^{i\Delta_o}$. From Lemma \ref{lemma}, we know that $e^{i\Delta}\in\sigma_p(U)$. Although, in the case  $e^{i\lambda}=e^{i\Delta _{o}}\ (\Delta _{o} \neq \Delta)$, (\ref{eq1}), (\ref{eq2}) become
\[
\begin{bmatrix}
\Psi _{1} (x-1)\\
\Psi _{3} (x)
\end{bmatrix} =\begin{cases}
T^{x-1}\begin{bmatrix}
k_{2}\\
k_{2}
\end{bmatrix} , & x >0,\\[15pt]
T^{x}\begin{bmatrix}
k_{1}\\
k_{1}
\end{bmatrix} , & x\leq 0,
\end{cases}  
\]
where $k_1,k_2\in\mathbb{C}$. 
By the similar discussion as Theorem \ref{Theorem Ker}, $e^{i\Delta_o}\in\sigma_p(U)$ if and only if followings hold:
\begin{align*}
    &1.\ \cos( \Delta_o -\Delta) -c>0,\\
    &2.\ \begin{bmatrix}1 \\1\end{bmatrix}\in\ker\left( T-\zeta ^{<}\right)\ or\   \begin{bmatrix}1 \\1\end{bmatrix}\in\ker\left( T -\zeta ^{>}\right).
\end{align*}
However, from (\ref{eq:ev1}), (\ref{eq:ev2}), we know  
$\begin{bmatrix}1 & 1\end{bmatrix}^{t}\notin\ker\left( T-\zeta ^{<}\right),\ \begin{bmatrix}1 & 1\end{bmatrix}^{t}\notin \ker\left( T -\zeta ^{>}\right)$ if $\cos( \Delta_o -\Delta) -c>0$, thus $\Psi\notin\mathcal{H}\setminus\{\mathbf{0}\}$ for all $k_1,k_2\in\mathbb{C}$ and $e^{i\Delta _{o}}\notin\sigma_p(U).$ Under the Assumption \ref{assumption}, i.e, $e^{i\lambda}\neq e^{i\Delta},e^{i\Delta_o}$, Theorem \ref{Theorem Ker} shows that $e^{i\lambda}\in\sigma_p(U)$ if and only if $ \cos( \lambda -\Delta ) -c >0$ and one of the
followings hold: 
\begin{align*}
1. &\ \sin^{2}( \lambda -\Delta _{o})(\cos( \lambda -\Delta ) -c) =( 1-\cos( \lambda -\Delta ))(\cos( \lambda -\Delta _{o}) -c_{o})^{2},\\
 and&\ \sin( \lambda -\Delta )\sin( \lambda -\Delta _{o})(\cos( \lambda -\Delta _{o}) -c_{o}) < 0.\\[7pt]
2. &\ ( 1-\cos( \lambda -\Delta _{o}))(\cos( \lambda -\Delta ) -c) =( 1-\cos( \lambda -\Delta ))( 1+\cos( \lambda -\Delta _{o})), \ \\
 and& \ \sin( \lambda -\Delta )\sin( \lambda -\Delta _{o})( 2\cos( \lambda -\Delta _{o}) +1-c_{o}) < 0.
\end{align*}
From these discussions, we have the following proposition:

\begin{proposition}
\label{pro:one-defect}
When $\Delta _{o} = \Delta$
\[
\sigma _{p}(U) =\begin{cases}
\left\{e^{i\Delta } ,e^{i\lambda_+},e^{i\lambda_-}\right\}, \  & c< c_{o},\\
\left\{e^{i\Delta }\right\}, \  & c\geq c_{o},
\end{cases}
\]
where 
\[e^{i\lambda_\pm}=\dfrac{c+c_{o}^{2} \pm i( 1+c_{o})\sqrt{1-c+2c_{o} -\left( c+c_{o}^{2}\right)}}{1-c+2c_{o}} e^{i\Delta }.\]
\end{proposition}
The examples of Proposition \ref{pro:one-defect}
 are shown in Figures \ref{fig:1}, \ref{fig:2}.

\subsection{Two-phase model}

Here, we consider the two-phase model, where $C_o=C_p$. In the case which $\lambda$ does not satisfy Assumption \ref{assumption}, i.e, $e^{i\lambda}= e^{i\Delta_m},e^{i\Delta_p}$, Lemma \ref{lemma} shows $e^{i\Delta_m},e^{i\Delta_p}\in\sigma_p(U)$. Next, under the Assumption \ref{assumption}, i.e., $e^{i\lambda}\neq e^{i\Delta_m},e^{i\Delta_p}$, Theorem \ref{Theorem Ker} shows that $e^{i\lambda}\in\sigma_p(U)$ if and only if
followings hold: 
\begin{align*}
1. &\ e^{i\lambda } =\frac{( c_{p} -c_{m}) \pm i\sqrt{2( 1-c_{m})( 1-c_{p})( 1-\cos( \Delta _{m} -\Delta _{p}))}}{\left(( 1-c_{m}) e^{-i\Delta _{p}} -( 1-c_{p}) e^{-i\Delta _{m}}\right)},\\
2. &\  \sin( \lambda -\Delta _{p})\sin( \lambda -\Delta _{m}) < 0,\\
3. &\  \cos( \lambda -\Delta _{m})  >c_{m}.
\end{align*}
Note that $\cos( \lambda -\Delta _{m})  >c_{m}$ is equivalent to $ \cos( \lambda -\Delta _{p})  >c_{p}$ in this model.
\begin{proposition}
\label{prop:two-phase1}
When $\Delta_m=\Delta_p=\Delta$
\[
\sigma _{p}( U) =\left\{e^{i\Delta }\right\}.
\]
\end{proposition}
\begin{proposition}
\label{prop:two-phase2}
When $c_m=c_p=c$ and $\Delta_m\neq\Delta_p$, let
\begin{align*}
condition\ 1: & \ \frac{\sin( \Delta _{m} -\Delta _{p})}{\sqrt{2( 1-\cos( \Delta _{m} -\Delta _{p}))}}  >-c,\\
condition\ 2: & \frac{\sin( \Delta _{m} -\Delta _{p})}{\sqrt{2( 1-\cos( \Delta _{m} -\Delta _{p}))}} < c.
\end{align*}

Then
\[
\sigma _{p}( U) =\begin{cases}
\left\{e^{i\Delta _{m}} ,e^{i\Delta _{p}}\right\} , & if\ neither\ condition\ 1\ nor\ condition\ 2\ holds,\\
\left\{e^{i\Delta _{m}} ,e^{i\Delta _{p}} ,e^{i\lambda }\right\} , & if\ condition\ 1\ holds\ and\ condition\ 2\ does\ not,\\
\left\{e^{i\Delta _{m}} ,e^{i\Delta _{p}} ,-e^{i\lambda }\right\} , & if\ condition\ 2\ holds\ and\ condition\ 1\ does\ not,\\
\left\{e^{i\Delta _{m}} ,e^{i\Delta _{p}} ,e^{i\lambda } ,-e^{i\lambda }\right\} , & if\ both\ conditions\ 1\ and\ 2\ hold,
\end{cases}
\]
where 
\[
e^{i\lambda } =\frac{i\left( e^{i\Delta _{p}} -e^{i\Delta _{m}}\right)}{\sqrt{2( 1-\cos( \Delta _{m} -\Delta _{p}))}}.
\]
\end{proposition}

\begin{figure}[H]
\begin{subfigure}[H]{0.49\textwidth}
\centering
\includegraphics[width=0.83\linewidth]{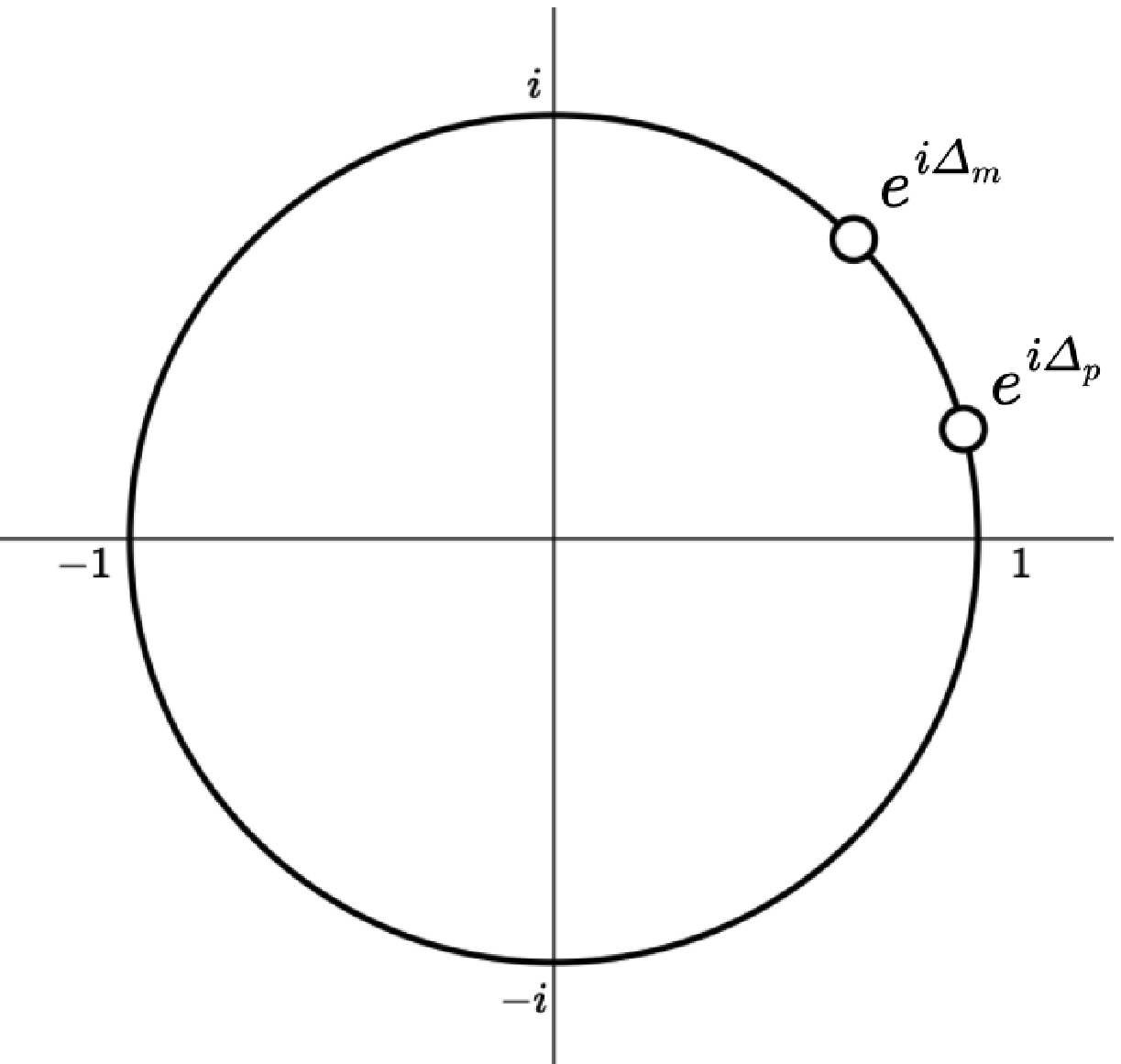} 
\caption{Eigenvalues of $U$}
\end{subfigure}
\begin{subfigure}[H]{0.49\textwidth}
\centering
\includegraphics[width=0.95\linewidth]{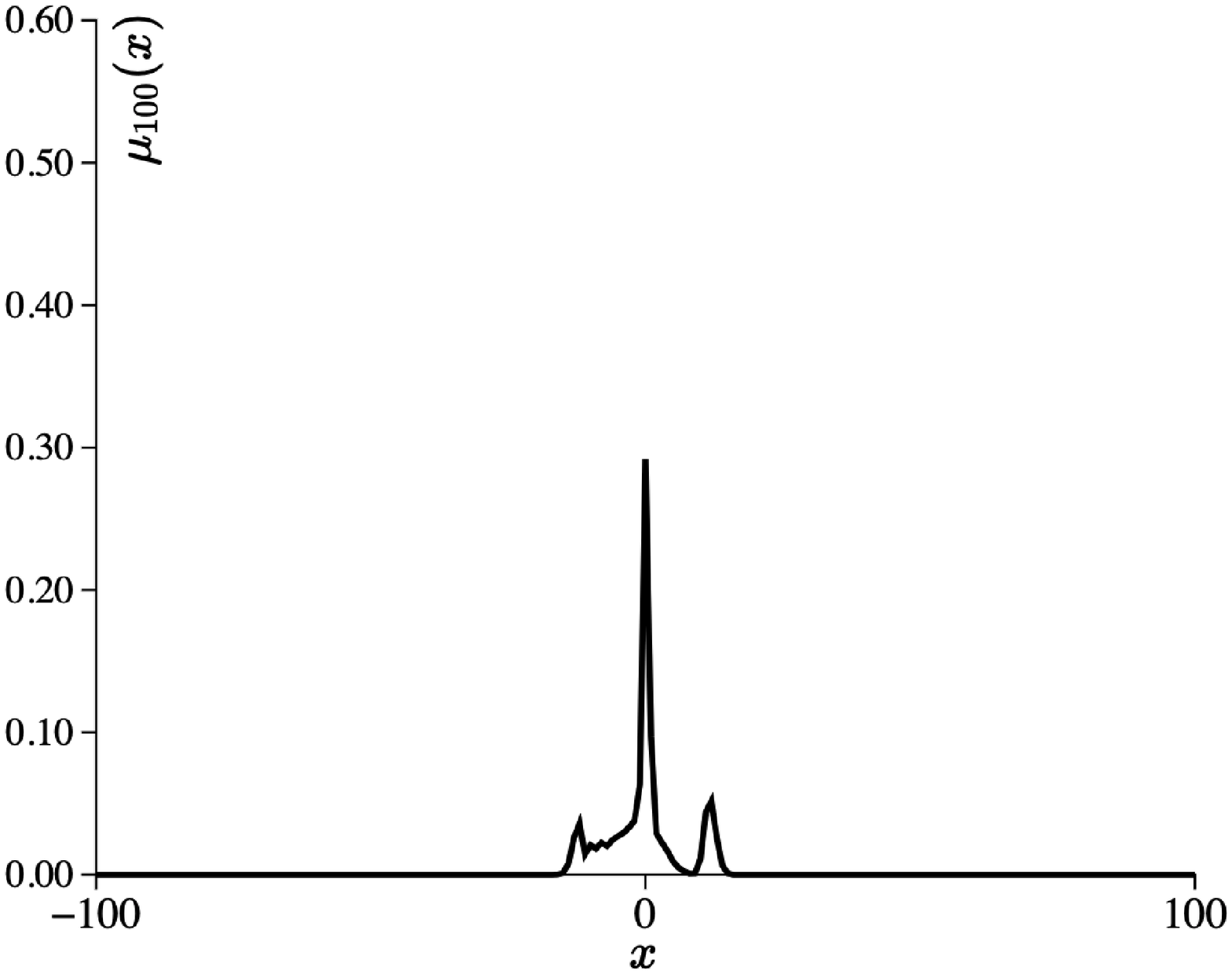}
\caption{Probability distribution}
\end{subfigure}
\caption{Example of Proposition \ref{prop:two-phase2} with parameters $\displaystyle \Delta _{m} =\frac{3}{12} \pi ,\ \Delta _{p} =\frac{1}{12} \pi ,\ \theta =\frac{11}{12} \pi $. (a) illustrates the eigenvalues, and (b) shows the probability distribution at time $100$ with initial state $\Psi_0(0)=[\frac{1}{\sqrt{3}},\frac{i}{\sqrt{3}},\frac{1}{\sqrt{3}}]^t$ and $\Psi_0(x)=\mathbf{0}$ for $x\neq 0$}
\label{fig:3}  
\end{figure}

\begin{figure}[H]
\begin{subfigure}[H]{0.49\textwidth}
\centering
\includegraphics[width=0.83\linewidth]{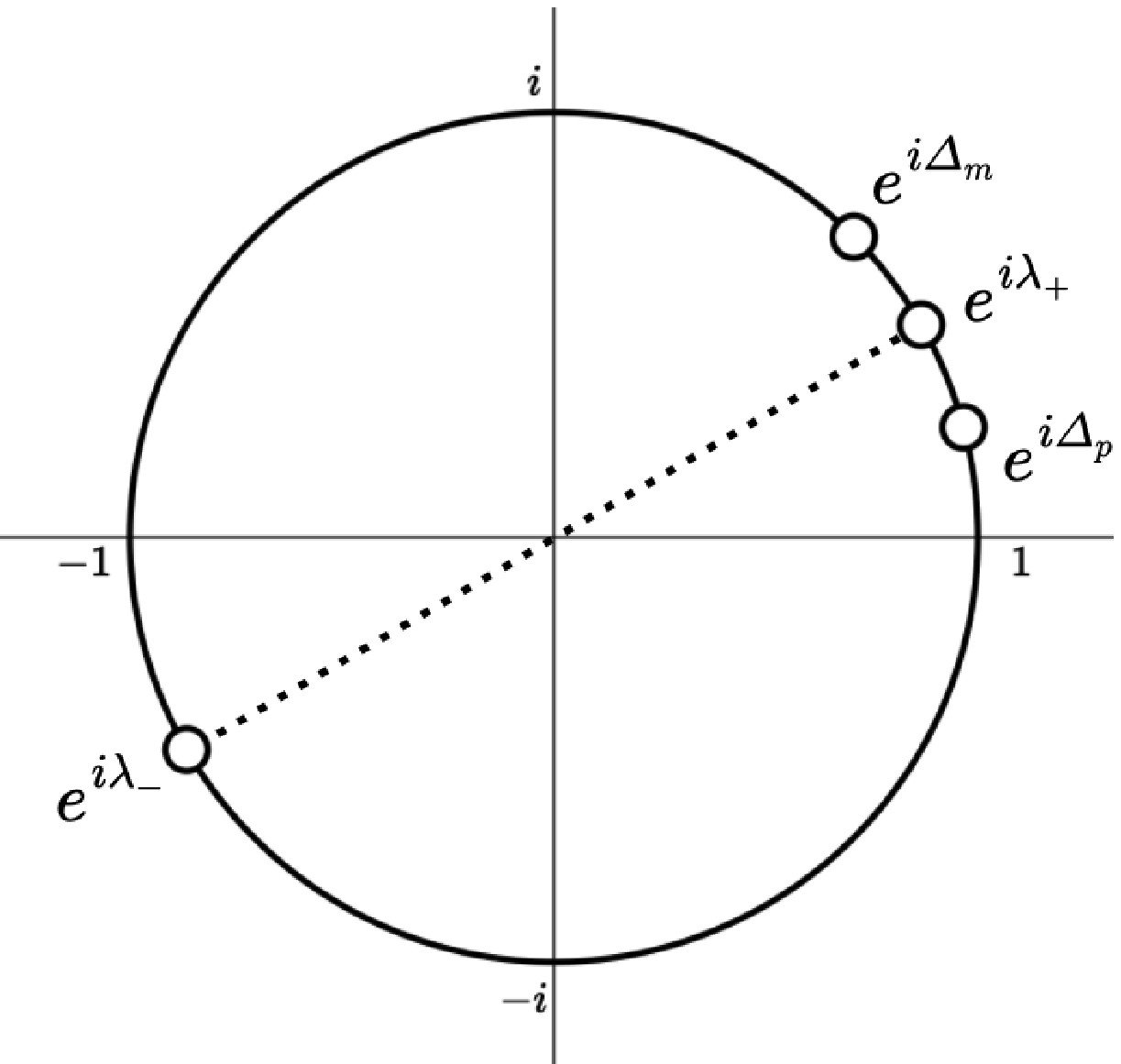} 
\caption{Eigenvalues of $U$}
\end{subfigure}
\begin{subfigure}[H]{0.49\textwidth}
\centering
\includegraphics[width=0.95\linewidth]{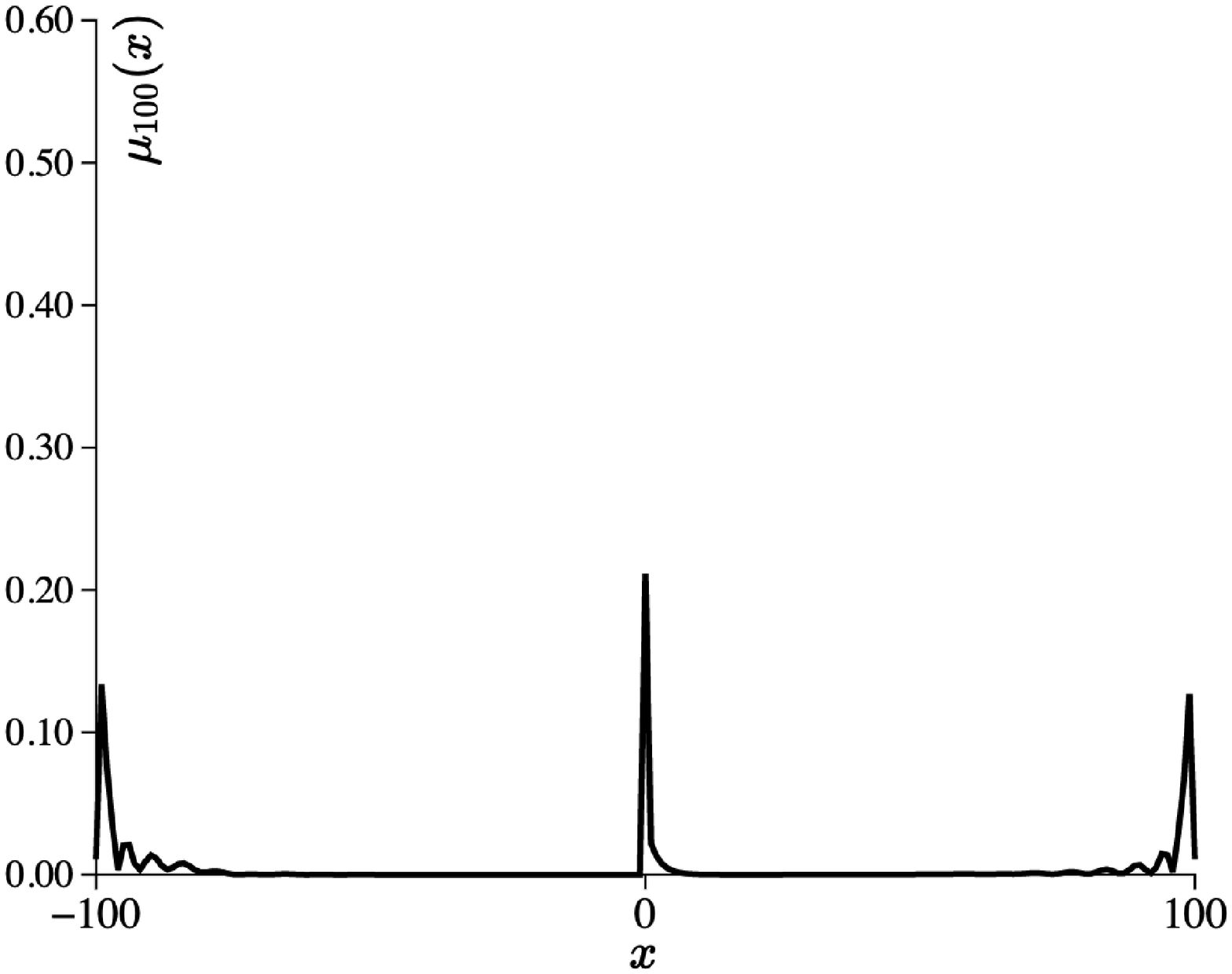}
\caption{Probability distribution}
\end{subfigure}
\caption{Example of Proposition \ref{prop:two-phase2} with parameters $\displaystyle \Delta _{m} =\frac{3}{12} \pi ,\ \Delta _{p} =\frac{1}{12} \pi ,\ \theta =\frac{1}{12} \pi $. (a) illustrates the eigenvalues, and (b) shows the probability distribution at time $100$ with initial state $\Psi_0(0)=[\frac{1}{\sqrt{3}},\frac{i}{\sqrt{3}},\frac{1}{\sqrt{3}}]^t$ and $\Psi_0(x)=\mathbf{0}$ for $x\neq 0$}
\label{fig:4}  
\end{figure}

The examples of proposition \ref{prop:two-phase2}
 are shown in Figures \ref{fig:3}, \ref{fig:4}.

\section{Summary}
In this paper, we analyzed eigenvalues of two-phase three-state generalized Grover walks with one defect in one dimension. In Section \ref{sec:2}, we successfully derived Theorem \ref{Theorem Ker} via transfer matrices, which is the necessary and sufficient condition for the eigenvalue problem for $n$-state QWs with $n-2$ self-loops under the Assumption \ref{assumption}. Next, we focused on the eigenvalue problem for three-state generalized Grover walks in Section \ref{sec:3}. Lemma \ref{lemma} revealed that $e^{i\Delta_{\pm\infty}}$ are eigenvalues of $U$, which also indicates that these models always exhibit localization. Subsequently, by applying Theorem \ref{Theorem Ker}, we got the necessary and sufficient condition for the eigenvalue problem and successfully calculated concrete eigenvalues for one-defect model in Propositions \ref{pro:one-defect}, and two-phase models in Propositions \ref{prop:two-phase1} and \ref{prop:two-phase2}. 

\section*{Acknowledgements}
The author expresses sincere thanks and gratitude to Kei Saito for helpful comments and discussion.
\printbibliography
\end{document}